\documentclass[a4paper,twoside]{article}
\newcommand{\affil}[1]{$^{\rm #1}$}
\usepackage[authoryear]{natbib}
\usepackage{pdflscape}
\usepackage{multirow}
\usepackage{longtable}

\bibpunct{(}{)}{;}{a}{}{,}
\usepackage{graphicx}
\usepackage{url}
\date{} 
\title{\large\bf\flushleft Scientific Visualization in Astronomy: Towards the Petascale Astronomy Era\thanks{Research undertaken as part of the Commonwealth Cosmology Initiative (CCI: www.thecci.org), an international collaboration supported by the Australian Research Council}}
\author{\parbox{\textwidth}{\flushleft
\vspace{-0.5cm}
{\it Amr Hassan\affil{A,B} and Christopher J.\ Fluke\affil{A}}\\
\vspace{0.4cm}
{\small \affil{A}\,Centre for Astrophysics and Supercomputing, Swinburne University of Technology, PO Box 218, Hawthorn, Australia, 3122}\\
{\small \affil{B}\,Email: ahassan@astro.swin.edu.au}}}
\begin{document}

\twocolumn[
\begin{changemargin}{.8cm}{.5cm}
\begin{minipage}{.9\textwidth}
\vspace{-1cm}
\maketitle

%
%
\small{\bf Abstract:} 
Astronomy is entering a new era of discovery, coincident with the establishment of new facilities for observation and simulation that will routinely generate petabytes of data. While an increasing reliance on automated data analysis is
anticipated, a critical role will remain for visualization-based knowledge
discovery.  We have investigated scientific visualization applications in astronomy through an examination of the literature published during the last two decades. We identify the two most active fields for progress -- visualization of large-$N$ particle data and spectral data cubes -- discuss open areas of research, and introduce a mapping between astronomical sources of data and data representations used in general purpose visualization tools. We discuss contributions using high performance computing architectures (e.g: distributed processing and GPUs), collaborative astronomy visualization, the use of workflow systems to store metadata about visualization parameters, and the use of advanced interaction devices.  We examine a number of issues that may be limiting the spread of scientific visualization research in astronomy and identify six grand challenges for scientific visualization research in the Petascale Astronomy Era.  


\medskip
\medskip
\end{minipage}
\end{changemargin}
]

\small


\section{Introduction}
Astronomy is a data-intensive science. Petabytes\footnote{1 petabyte = $10^{15}$ bytes} of observational data is already in stored archives \citep{Brunner:2002, szalay:2001}, even before facilities such as the Atacama Large Millimeter Array [ALMA; \citet{brown:2004}], 
the Large Synoptic Survey Telescope [LSST; \citet{Ivezic:2008}], 
LOFAR \citep{rottgering:2003}, SkyMapper \citep{keller:2007}, the Australian Square Kilometre Array Pathfinder [ASKAP; \citet{johnston:2008}],the Karoo Array Telescope [MeerKAT; \citep{booth:2009}], and ultimately the Square Kilometre Array itself, reach full operational status.  
Cosmological simulations with $10^{10}$ particles [e.g. \citet{springel:2005,Klypin:2010}] are also producing many-terabyte datasets, and the highest resolution simulations codes executed on the next generation of petaflop/s supercomputers will result in further petabytes of data.  
The Petascale Astronomy Era is a natural outcome of current and future major observatories and supercomputer facilities.  

Astronomy sits alongside fields such as high-energy physics and bioinformatics in terms of the data volumes that are available to its practitioners.
Such data volumes pose significant challenges for data analysis, data storage and access, leading to the development of a fourth 
data-intensive (or eScience) paradigm for science \citep{szalay:2006,bell:2009}. Much work will be required to find effective solutions for knowledge discovery in the Petascale Astronomy Era, with a likely emphasis on automated analysis and data mining processes \citep{ball:2009,Borne:2009,pesenson:2010}.
However, a critical step in understanding, interpreting, and verifying the outcome of automated approaches requires human intervention. This is most easily achieved by simply looking at the data: the human visual system has powerful pattern recognition capabilities that computers are far from being able to replicate.

Astronomy has a long history in the use of diagrams, maps and graphs\footnote{See \citet{funkhouser:1936} for an account of the earliest extant 
astronomical graph.} to explain concepts, aid understanding, present results, and to engage the public.  However, there is more to {\em visualization} than just making pretty pictures.  Data visualization is a fundamental, enabling technology for knowledge discovery, and an important research 
field in its own right.    

The broader field of astronomy visualization encompasses topics such as optical and radio imaging, presentation of simulation results, 
multi-dimensional exploration of catalogues, and public outreach visuals.  
Aspects of visualization are utilized in the various stages of astronomical research - from the planning stage, through the observing process or
running of a simulation, quality control, qualitative knowledge discovery 
and quantitative 
analysis.\footnote{These are distinct phases - see \citet{djorgovski:2005}.}  Indeed, much of astronomy deals with the process of making and
displaying two-dimensional (2D) images (e.g. from CCDs) or graphs which 
are suitable for 
publication in books, journals, conference presentations and in education [see
\citet{fluke:2009} for alternatives]. 

An important sub-field of visualization is {\em scientific visualization}: 
the process of turning (numerical) data with dimensionality $N \geq 3$, 
usually with an inherent geometrical structure, into images that 
can be inspected by eye.  At its conception in the 1980s \citep{mccormick:1987,frenkel:1988,defanti:1989}, scientific visualization was 
envisaged as an interactive process, with an emphasis on understanding 
and analysis of data (including qualitative, comparative and
quantitative stages), not just presentation (Wright 2005). 
Research in this field includes techniques for displaying data (e.g. 
through the use of surface rendering, volume rendering, streamlines, etc.), 
efficient implementations of display algorithms for 
increasingly complex data and data structures (including both data 
dimensionality and dataset size) while retaining interactivity, and 
effective use of high-performance computing for tasks such as
parallel rendering and computational steering (where interaction 
with a simulation occurs during processing, and helps to drive 
the direction of the next stage of processing).
We refer the interested reader to the general introductions by 
\citet{gallagher:1995}, \citet{johnson:2004}, and \citet{schroeder:2006}. 

There is a very subtle distinction between scientific visualization and the 
closely-related field of information visualization \citep{spence:2001}.
The latter deals with presentation and understanding of multi-dimensional 
data, where the search for relationships between data points 
is the motivation for investigation.  The following example attempts to 
highlight 
the difference:  a map of the locations of normal elliptical galaxies with a colour scheme 
or symbols relating to mean surface brightness, effective surface brightness and
central velocity disperson (so that the emphasis is on the spatial arrangment) 
is a scientific visualization; a three-dimensional plot of 
these last three quantitites demonstrating how they form the fundamental plane
\citep{djorgovski:1987} is an information visualization.  Similarly a 
three-dimensional (3D) plot
of the $(x,y,z)$ locations of particles from an $N$-body simulation, where no
colour coding is used to present additional numerical properties is
information visualization (essentially a 3D scatter plot),
but colouring particles by local density or temperature, or the use of a surface
or volume rendering technique to identify large-scale structures, is 
scientific visualization.   We make use of this distinction in order to help 
identify research work that is relevant for our overview, and for brevity
use ``visualization'' hereafter to mean three-dimensional scientific visualization. 

As a multidisciplinary field, scientific visualization has been used 
with great success in medical imaging, molecular modelling, engineering
(e.g. computational fluid dynamics), architecture, and astronomy. 
Scientific visualization of astronomical data includes both observational
data generated over a variety of wavelengths (optical, radio, 
X-ray, etc.) or from computer simulations.  Visualization of 
observational data poses some specific challenges in terms of the data 
volume, dynamic range, (often low) signal-to-noise ratio, incomplete or 
sparse sampling, and astronomy-specific coordinate systems. 
For simulations, challenges include the number of particles, 
mesh resolution, and range of length and time scales. While any one of these issues 
is not unique to astronomy, taken as a whole, effective astrophysical 
visualization requires its own unique solutions.  

One of the first systematic 
astronomy visualization trials was undertaken by Gitta Domik, 
Kristina Mickus-Miceli, and collaborators
at the University of Colorado [\citet{mickus:1990a,mickus:1990b,domik:1992a,
domik:1992b,brugel:1993}].  They developed a prototype application 
named the Scientific Toolkit for Astrophysical Research (STAR) using IDL  on 
top of X-Windows. Their main goals were to offer visualization tools that 
were driven by the needs of astronomers, and that would integrate 
with existing data analysis tools.  STAR's main functionality included 
display of one and two-dimensional datasets, perspective projection, 
pseudo-colouring, interactive colour coding techniques,
volumetric data displays, and data slicing. STAR was introduced as a 
prototype to prove the feasibility of the user interface, and 
visualization techniques proposed in their report. 

\citet{norris:1994} presented a blueprint for visualization research
in astronomy,  highlighting the suitability of 3D visualization
for providing an intuitive understanding that was missing when 
using 2D approaches (e.g. data slicing, where 
individual channels are examined separately or played back as movie, 
requiring the viewer to remember what was seen in earlier channels).  
Visualization techniques could enable features of the data to be seen 
that would otherwise have 
remained unnoticed, such as low signal-to-noise structures extending 
over multiple channels.  \citet{norris:1994} noted the importance
of visualization in communicating results qualitatively, but 
identified quantitative visualization as the missing ingredient 
that would allow true interactive hypothesis testing - an essential 
part of the scientific process.  While still relevant today, several key aspects 
-- most notably a wider uptake of 3D visualization by astronomers 
-- have yet to be fully realised.  

\subsection{Scope and Purpose}
We consider the development, advancement and application 
of scientific visualization techniques in astronomy over the last two decades - 
coincident with the lifetime of scientific visualization as a field 
of inquiry in its own right.   
To our knowledge, there have been no previous attempts to examine the status 
of scientific visualization in astronomy.  The Masters thesis by 
\citet{Horacio:2003} discusses visualization strategies for several
numerical datasets from astronomical simulations; \citet{leech:2005} 
surveyed visualization software available for radio and sub-mm data; 
\citet{dubinski:2008} provided an introduction to particle visualization 
as a companion to a brief history of $N$-body methods;  
\citet{kapferer:2008} considered software and hardware 
visualization requirements for numerical simulations;
\citet{li:2008} described strategies for dealing with multiwavelength data;
and \citet{fluke:2009} summarised some basic elements of 
cosmological visualization for observational and simulation data.  

It is not our intent to provide a comprehensive account of all research 
outcomes that have made use of scientific visualization tools or 
approaches [see \citet{brodbeck:1998}, \citet{Hultquist:2003},
or the growing scientific output from the AstroMed project,
\citet{borkin:2008}, \citet{goodman:2009}, and \citet{arce:2010} for
representative examples], but to investigate how scientific
visualization in astronomy has advanced over the last two decades.  
In particular, we do not consider geographical-style 3D visualization from 
planetary missions, or data from solar physics [see 
\citet{ireland:2009}, and articles therein, or \citet{aschwanden:2010} 
for a complementary review of the latter field].
We aim to provide an overview for researchers who wish to understand the state of scientific visualization
in astronomy, with an emphasis on simulation and spectral cube data, aiming to 
motivate future work in this field. We assert that many current astronomy 
visualization approaches and applications are incompatible with the 
Petascale Astronomy Era, and much work is required to ensure that astronomers 
have the tools they need for knowledge discovery over the next decade and beyond.  

The remainder of this paper is set out as follows.  In section 
\ref{sct:review} we present an overview of progress in scientific 
visualization in astronomy, from projects in the early 1990s until the present. 
We pay special attention to two important classes of data: 
large-$N$ particle systems (section \ref{sct:largeN}), and spectral data cubes (section \ref{sct:SDC}).
We consider other research areas including distributed (section \ref{sct:dv}) and collaborative visualization (section \ref{sct:cv}), image display 
(section \ref{sct:image}), workflow (section \ref{sct:workflow}),
 and public outreach visuals 
(section \ref{sct:public}).  In section \ref{sct:software}, we demonstrate how the nature of astronomical data impacts on the choice
of visualization software, highlighting some of the advantages 
and disadvantages of using general visualization packages instead of
custom astronomy code.  
In section \ref{sct:discuss}, we discuss some of the challenges that 
astrophysical visualization must overcome in order to be useful 
and usable, and identify six grand challenges for scientific visualization
research in the Petascale Astronomy Era.
Finally, we present our concluding remarks in section \ref{sct:conclusion}.  

For a article on visualization, it may seem surprising that so few images have been included. Our preference is for the reader to view 
the original, published versions, appearing as their author(s) intended, rather than attempting to replicate them here. Any omission of significant work or software related to scientific visualization in astronomy is wholly the 
responsibility of the authors.  
\section{Scientific Visualization in Astronomy}
\label{sct:review}

\subsection{Visualization Techniques}
\label{sct:outline}

\begin{figure*}
\begin{centering}
\includegraphics[width=6.2in]{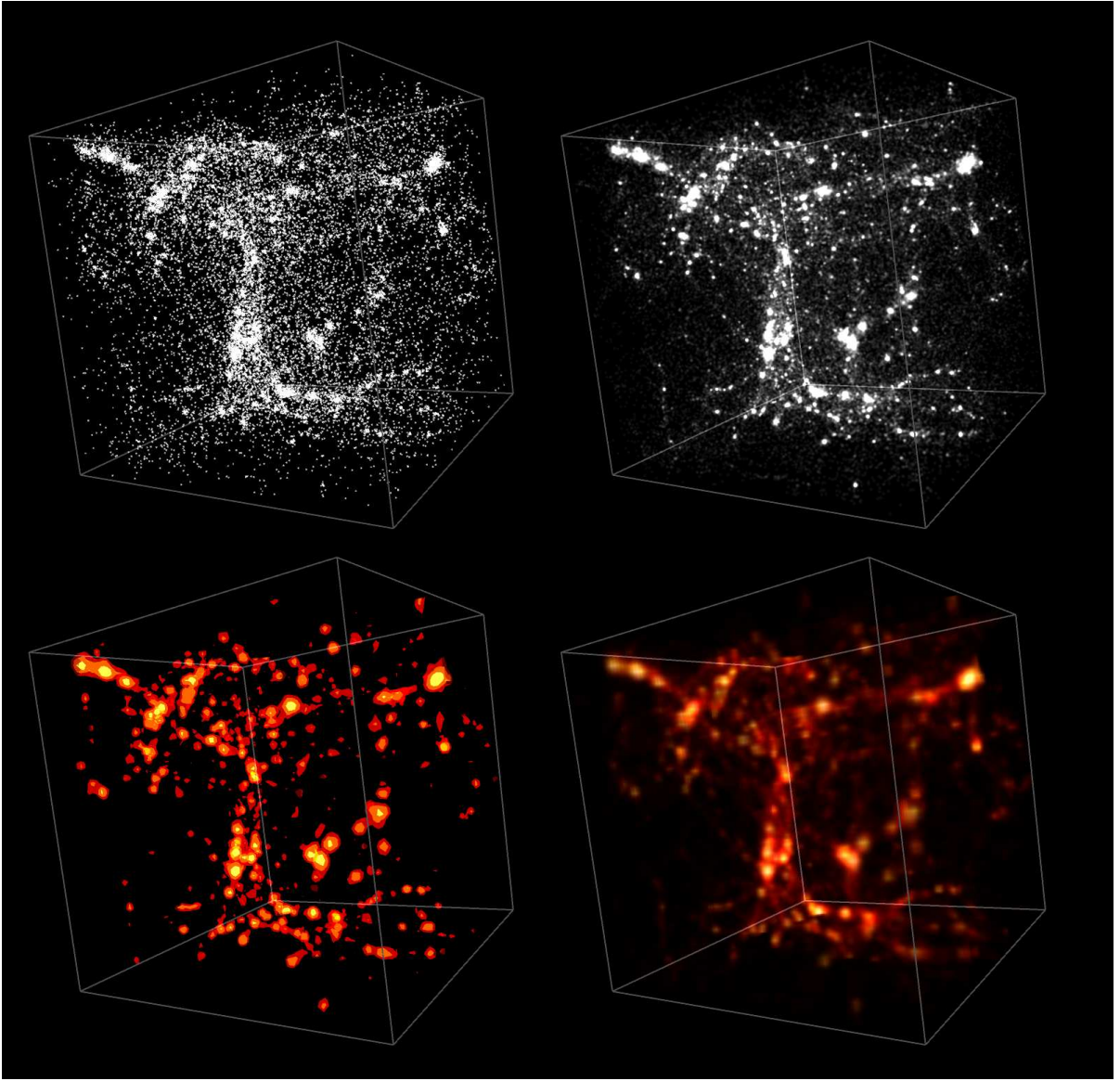}
\caption{Four common visualization methods applied to a cosmological $N$-body
dataset.
Scattered point data (top left); Gaussian ``splats'', using
transparent-mode texture blending (top right);
three representative isosurface levels with density increasing
from red-orange-yellow (bottom left); and texture-based volume rendering with
``heat'' colour map increasing from black through red and yellow to white (bottom right).
Data courtesy Madhura Killedar (University of Sydney). Visualization
was performed using S2PLOT.}
\label{fig:compare}
\end{centering}
\end{figure*}

As a starting point to exploring advancements in scientific visualization in 
astronomy, we introduce the most common techniques for presenting three-dimensional scalar data in astronomy: points, splats, isosurfaces and volume rendering.\footnote{We found few
papers that explicitly discussed the use of streamlines as a visualization
technique. Outside of solar astronomy, these are more commonly used to
understand flows in, e.g. computational fluid design or geophysics
visualizations.}  Figure \ref{fig:compare} shows how the same dataset, in
this case a single snapshot from a cosmological simulation, appears when 
rendered using the four techniques.

Plotting points (top left) as fixed width pixels is often the most straightforward
representation, however, this approach is limited by the available 
resolution (or pixel density) of the display. Splatting (top right) uses
small textures, often with a Gaussian intensity profile, to replace 
point-like objects.  Splats are billboards, i.e. they always point towards the
virtual camera regardless of the orientation of the scene, and scale better 
with distance than pixels.  Combining splats on the graphics card gives an
effect like volume-rendering, but without the calculation overhead of ray-tracing.

An isosurface (bottom left) or isodensity surface is a three-dimensional
equivalent of contouring. Common methods for calculating an isosurface
from a dataset include marching cubes \citep{lorensen:1987,montani:1994}, marching tetrahedra \citep{bloomenthal:1994}, multiresolution isosurface extraction \citep{gerstner:2000}, and surface wavefront propagation \citep{wood:2000}. Isosurfaces are usually used to search for correlation between different scalar variables, but are less useful to give a global picture of the dataset. Volume rendering (bottom right) attempts to provide a global view of the dataset, particularly useful to render both the external surfaces and the interior 3D structures with the ability to display weak or fuzzy surfaces. Volume rendering can be performed using ray-tracing or using the graphics 
card to combine a series of (semi-)transparent texture maps [e.g. \citet{cabral:1994}, this approach was used for 
Figure \ref{fig:compare}].

\begin{figure*}
\begin{centering}
\includegraphics[width=6.2in]{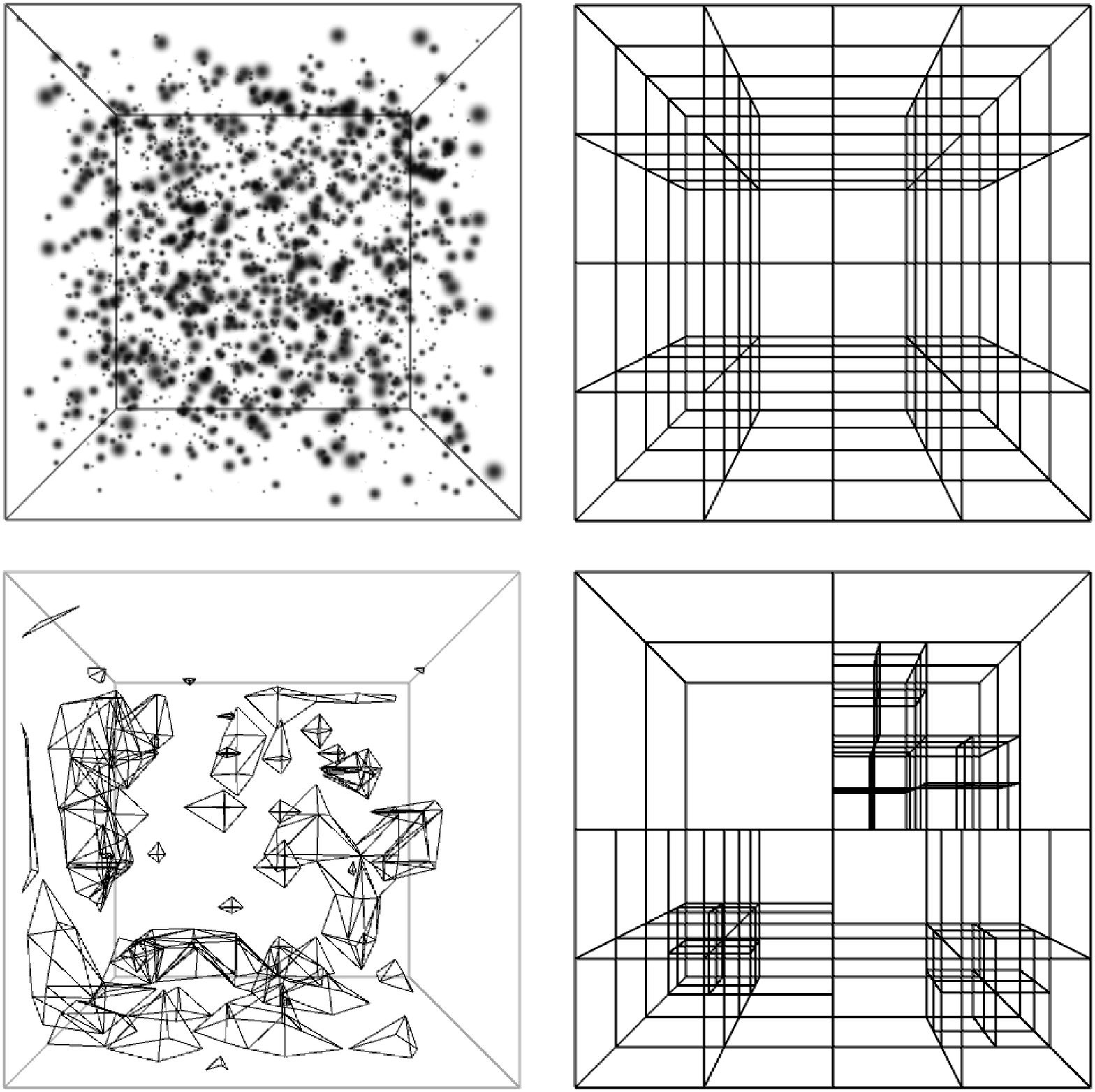}
\caption{The four standard data representations used in scientific
visualization. 
Scattered point data (top left); structured grid (top right); unstructured grid (bottom left); and adaptive/multi-resolution
grid (bottom right). Visualization was performed using S2PLOT. Note that the bounding box surrounding the unstructured grid is for reference only.}
\label{fig:datatypes}
\end{centering}
\end{figure*}

\subsection{The Nature of Astronomical Data}
\label{sct:DataNature}
\begin{figure*}
\begin{center}
\includegraphics[scale=0.45]{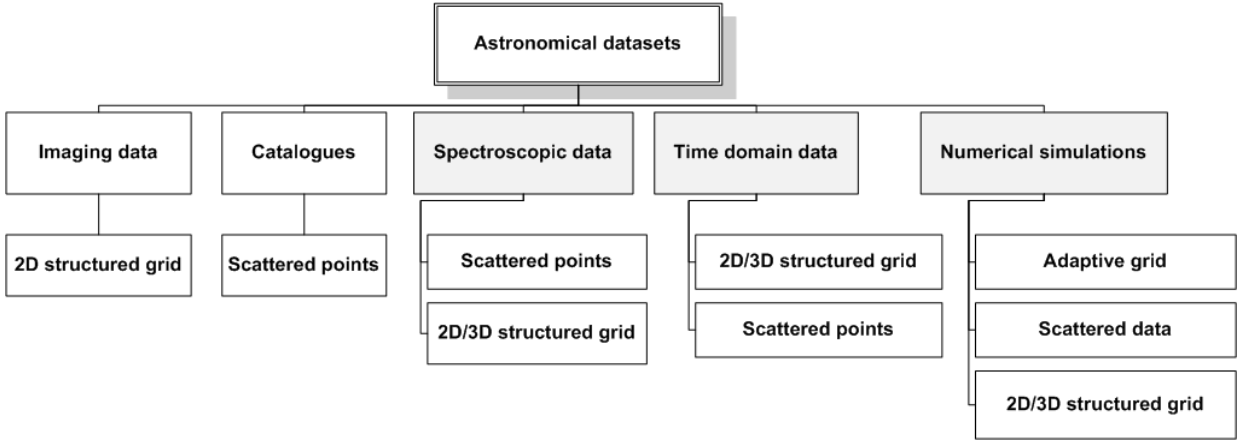}
\caption{The nature of astronomical data, demonstrating a mapping between
sources of data \citep{Brunner:2002} and data 
representations \citep{gallagher:1995}.  We have selected the most common
representations of data for each of the sources.}
\label{fig:classification}                                
\end{center}                                 
\end{figure*}

The nature of data has an impact on the choice of visualization technique, and 
hence software. 
One way to look at astronomical data \citep{Brunner:2002} is to consider the
origin or physical source:
\begin{itemize}
\item {\em Imaging data}: two-dimensional within a narrow wavelength range at a particular epoch.
\item {\em Catalogues}: secondary parameters determined from processing of image data (coordinates, fluxes, sizes, etc). 
\item {\em Spectroscopic data and associated products}: this includes 
one-dimensional spectra and 3D spectral data cubes, data on distances obtained from
redshifts, chemical composition of sources, etc.
\item {\em Studies in the time domain}: including observations of moving objects, variable and transient sources which require multiple
observations at different epochs, or synoptic surveys.
\item {\em Numerical simulations from theory}: which can include properties such as spatial position, velocity, mass, density, temperature, and particle type. These
properties may also be presented with an explicit time dependence through the
use of ``snapshot'' outputs.
\end{itemize}
An alternative classification [see \citet{gallagher:1995}] is based on how 
the data is representated programmatically, i.e. how it is stored and organized 
in memory or on disk:
\begin{itemize}
\item {\em Scattered points}: data is comprised of a set of point locations $(x,y,z)$ and associated data attributes (e.g. density, pressure, and temperature).
\item {\em Structured grid}: data values are specified on a regular 
three-dimensional grid, with grid cells aligned with the Cartesian axes.
\item {\em Unstructured grid}: data values are specified on corners of a 2D/3D shape element with an explicitly defined connectivity.
\item {\em Adaptive grid}: data values are specified on a multi-resolution structured grid. A coarse grid is used to cover the entire computational domain combined with superimposed sub-grids to provide higher resolution for regions of interest (e.g. where particle density is highest).
\end{itemize}
Figure \ref{fig:datatypes} demonstrates each of these representations.
This second classification is more familiar to practioners of scientific 
data visualization, and is one that helps guide the choice of 
visualization techniques in a way that is somewhat independent of a particular
scientific domain. 
Figure \ref{fig:classification} demonstrates, in broad terms, how the
astronomically-motivated data categories can be mapped into the 
data representation schema. 
Unstructured-grids are rarely used in astronomy because they are usually generated from finite element analysis or domain decomposition methods, which are not widely used in astronomy [see \citet{Springel:2010} for an example of unstructured grid usage]. 

With regard to 3D scientific visualization, we are primarily interested 
in spectroscopic, time domain-data, and numerical simulations, but there may 
be a need to overlay images (e.g. image slicing), or secondary catalogue data (e.g. for comparisons 
between multiwavelength data).  
Based on the astronomical data category, the majority of visualization papers 
that we identified related to either $N$-body and other large-scale
particle simulations or spectral data cubes. 
We now summarize the main developments in each of these two areas. 
In particular,
we focus on the implementation trends, as this provides a useful starting
point for future work in this field, and occurences of the visualization 
techniques outlined in Section \ref{sct:outline}.  We do not attempt to provide
specific technical details of each visualization solution, as some approaches
are now out-dated due to improvements in graphics hardware -- most notably 
through the appearance of low-cost, massively parallel graphics 
procesing units (GPUs).

\begin{table*}
\caption{The classification of contributions dedicated to the visualization 
of large-$N$ particle simulations. See Section \ref{sct:outline} and \ref{sct:DataNature} for a description of the visualization techniques and dataset types. The methodology column highlights the main implementation trend used in each work.  }
\label{tab:ListofNParticlePapers}
\centering
\begin{tabular}{|p{4.5cm}|c|c|c|c|c|c|c|}
\hline
& \multicolumn{3}{|c|}{Dataset Type} & \multirow{2}{*}{Methodology} & \multicolumn{3}{|c|}{Visualization Techniques}\\ \cline{2-4} \cline{6-8}
&AMR&Scattered&Structured&&Isosurface&VR&Points\\ \hline
\citet{ostriker:1997}&$\bullet$&&&Custom&&$\bullet$&$\bullet$\\ \hline
\citet{norman:1999}&$\bullet$&&&Custom&&$\bullet$&$\bullet$\\ \hline
\citet{becciani:2000, becciani:2001, gheller:2002, gheller:2003, amati:2003}&&$\bullet$&$\bullet$&Modify&$\bullet$&$\bullet$&$\bullet$\\ \hline
\citet{teuben:2001}&&$\bullet$&&Modify&&&$\bullet$ \\ \hline
\citet{kahler:2002,kahler:2002IEEE,kahler:2003,kaehler:2006}&$\bullet$&&&Modify&&$\bullet$&\\ \hline
\citet{Horacio:2003}&&&$\bullet$&Existing&&$\bullet$&\\ \hline
\citet{hopf:2003,hopf:2004}&&$\bullet$&&Custom&&$\bullet$&\\ \hline
\citet{staff:2004}&&&$\bullet$&Modify&$\bullet$&&$\bullet$ \\ \hline
\citet{ahrens:2006}&$\bullet$&&&Modify&$\bullet$&$\bullet$&$\bullet$\\ \hline
\citet{navratil:2007}&&$\bullet$&$\bullet$&Existing&$\bullet$&&$\bullet$\\ \hline
\citet{kahler:2007}&&$\bullet$&&Custom&&$\bullet$&\\ \hline
\citet{comparato:2007}&&$\bullet$&$\bullet$&Modify&$\bullet$&$\bullet$&$\bullet$\\ \hline
\citet{biddiscombe:2007}&&$\bullet$&&Modify&&&$\bullet$  \\ \hline
\citet{linsen:2008}&&$\bullet$&&Custom&$\bullet$&$\bullet$&\\ \hline
\citet{dubinski:2008}&&$\bullet$&&Custom&&$\bullet$&\\ \hline
\citet{Szalay:2008}&&$\bullet$&&Custom&&$\bullet$&\\ \hline
\citet{splotch:2008}&&$\bullet$&&Custom&&$\bullet$&\\ \hline
\citet{kapferer:2008}&&$\bullet$&$\bullet$&Existing&$\bullet$&&$\bullet$\\ \hline
\citet{nakasone:2009}&&$\bullet$&&Custom&&&$\bullet$\\ \hline			
\citet{farr:2009}&&$\bullet$&&Custom&&&$\bullet$\\ \hline
\citet{Splotch:2010}&&$\bullet$&&Custom&&$\bullet$&\\ \hline
\citet{becciani:2010}&&$\bullet$&$\bullet$&Modify&$\bullet$&$\bullet$&$\bullet$\\ \hline			
\end{tabular}
\end{table*}

\subsection{Large-$N$ Particle Simulations}
\label{sct:largeN}

For a science where direct experimentation is challenging, numerical simulations provide astronomers with a link between observations and theory.  Continued growth in processing power, new architectures and improved algorithms, have all enabled simulations to increase in resolution and accuracy.
Most astronomy simulations use of one of three main data representations (see Figure \ref{fig:classification}):
\begin{enumerate}
	\item Multi-dimensional scattered data: such as the GADGET-2 \citep{springel:2005} file format. Here, each data point is characterised by 
a set of spatial coordinates, with additional scalar and vector properties. 
This can be mapped to a regular scattered data visualization problem.  
	\item	3D/2D structured grid: where the point-data is distributed over a regularly-spaced mesh with a predefined resolution using smoothing techniques such as cloud-in-cell \citep{hockney:1988}. This data representation may cause a loss in fine details on scales below the mesh-size \citep{hopf:2004}.  
	\item	 Adaptive/multi-resolution grid: where the point-data is distributed over an adaptive mesh with a variable resolution. This is often the best data representation to cover a wide range of spatial and temporal domain with a minimal data size \citep{kahler:2002IEEE}. However, implementation and handling of boundary conditions between scales can be challenging.
\end{enumerate}

Table \ref{tab:ListofNParticlePapers} summarizes contributions 
dedicated to the $N$-particle visualization  problem with a 
classification based on the dataset representation type, the 
implementation trend, and the main visualization techniques used. 
It is noteworthy here that \citet{Horacio:2003}, \citet{staff:2004}, 
\citet{biddiscombe:2007}, and \citet{kapferer:2008}  presented the usage of vector plots as one of the visualization techniques.  
Although it is not commonly used to represent vector quantities from simulations in astronomy, vector visualization is a very useful 
technique as has been demonstrated in other fields, especially computational fluid dynamics. However, interpreting the physical meaning of vector fields is a higher-level cognitive task compared to identifying structures through the use of surface or volume rendering.

Visualizing $N$-particle data is a problem common to many scientific fields (e.g. point-based surface representations for 3D geometry processing \footnote{see \citet{kobbelt:2004} for a discussion about different available point-based rendering techniques}), and this has been an area for active research over the last two decades.  Such datasets are most ammenable to the use of general purpose scientific visualization tools/libraries, although there is still a tendency for astronomers to develop their own solutions.  In the astronomical literature, we identify three main approaches to implementation.

The first approach is to use an existing general purpose tool as-is \citep{Horacio:2003, navratil:2007, kapferer:2008}, however a data preprocessing step is usually required to convert the data into a suitable representation before the visualization is performed. \citet{Horacio:2003} used IDL to visualize 2D/3D numerical simulations of magneto-hydrodynamic clouds in the interstellar medium, stellar jets from variable sources, neutron star - black hole coalescence, and accretion disks around a black hole. \citet{navratil:2007} used Paraview and Partiview to render a time depended dataset of the first stars and their impact on cosmic history. \citet{kapferer:2008} presented the usage of IFrIT, MayaVi data visualize, Paraview, and VisIT with a discussion of the visual quality aspects, generating interactive 3D movies, real-time vector field visualization, and high-resolution display techniques. They also showed the usage of VisTrials workflow system for saving visualization meta-data. 

The second approach was to modify or extend existing general purpose tools  \citep{becciani:2000,becciani:2001,gheller:2002,kahler:2002IEEE, gheller:2003,amati:2003,ahrens:2006,becciani:2010}. \citet{becciani:2000,becciani:2001}, \citet{gheller:2002,gheller:2003}, and \citet{amati:2003} described the development of the AstroMD tool (a tool developed within the European Cosmo.Lab project\footnote{http://cosmolab.cineca.it/}). Its user interface was built based on TCL/TK while its visualization functionality uses VTK. It includes visualization capabilities such as:	isosurfaces, volume rendering, point picker, and sphere sampler.This work was continued by \citet{comparato:2007}, and \citet{becciani:2010} through the VisIVO project. VisIVO was developed based on the Multimod application framework (MAF). \citet{kahler:2002}, \citet{kahler:2002IEEE}, and \citet{kahler:2003,kaehler:2006} introduced an AMIRA\footnote{http://www.amira.com/} extension to render adaptive mesh refinement datasets. Their work was initiated within the framework of a television production for the Discovery Channel for rendering for the first stars in the universe.    

The final approach is to develop a custom system or library from scratch. From Table \ref{tab:ListofNParticlePapers}, we can say it is quite a popular choice \citep{ ostriker:1997, norman:1999, kahler:2007, linsen:2008, dubinski:2008, Szalay:2008, nakasone:2009, Splotch:2010}.  \citet{ ostriker:1997}, and \citet{norman:1999} described the work done within the Computational Cosmology Observatory (CCO) which act as an environment analogous to an astronomical observatory. Its implementation included: a specialized I/O library to handle HDF files, a desktop visualization tool, virtual-reality navigation and animation techniques, and Web-based workbenches for handling and exploring AMR data.  

\citet{kaehler:2006} and \citet{kahler:2007} used a GPU-assisted ray casting algorithm to provide a high quality volume rendering of AMR datasets. They avoided re-sampling the point's data onto a structured grid by directly encoding the point data in a GPU-octree data structure. \citet{linsen:2008} adopted a visualization approach based on isosurface extraction from multi-field particle volume data. They projected the $N$-dimensional data into  3D star coordinates to help the user to select a cluster of features. Based on the segmentation property induced by the cluster membership, a surface is extracted from the volume data. \citet{dubinski:2008} presented the  MYRIAD library . MYRIAD has been integrated with two different parallel N-Body codes (PARTREE\footnote{http://www.sdsc.edu/pub/envision/v15.2/hernquist.html} and GOTPM\footnote{http://www.cita.utoronto.ca/~dubinski/gotpm/}).

\citet{Szalay:2008} implemented a system that uses hierarchical level-of-details (LOD) for particle-like cosmological simulations, in order to display accurate results without loading in the full dataset. They were able to achieve a framerate of 10 frame per second with a desktop workstation and NVidia GeForce 8800 graphics card.  

The last noteworthy work in this direction is that done by \citet{splotch:2008} and \citet{Splotch:2010}. They introduced a tool to render directly point-like data in the GADGET-2 format. They used ray tracing to render in a fast and effective way the different families of point-like data. The same algorithm was enhanced to use GPUs with CUDA and distributed clusters using MPI. 

There is no clear choice as to which approach should be favoured, as there is a strong dependence on the visualization objectives and target. 
Most of the users of the first and the second trends aim to minimize the development cost and to use existing, tested, and open source packages with minimal or no modification. On the other hand, the researchers using the third approach aimed to enable the use of available advanced or specialized 
hardware infrastructure [e.g. the Grid environment \citep{becciani:2010}, or GPUs \citep{kaehler:2006} and \citep{Splotch:2010}]; produce better or faster
visualization results through customizing an existing visualization technique [\citet{hopf:2004,splotch:2008,Szalay:2008}]; or utilizing new platforms, such as the web-platform or virtual environments, and provide the users with better user interfaces or support collaborative interaction [\citet{nakasone:2009} and \citet{becciani:2010}].

Visualization approaches and software have needed to keep pace with improvements in simulation techniques and resolution (which can include
an increase in $N$ or the number of grid cells).  Indeed, visualizing ``large-$N$'' datasets, relative to the era of implementation,  
was addressed by most of the works [\citet{ostriker:1997,norman:1999,welling:2000,kaehler:2006,kahler:2007,Szalay:2008,Splotch:2010,becciani:2010}]. Attempts to solve this problem included the use of Grid-computing or a distributed 
cluster as the computing infrastructure \citep{ostriker:1997, norman:1999, Splotch:2010, becciani:2010};
using GPUs as the computing infrastructure \citep{ahrens:2006,kaehler:2006,biddiscombe:2007,kahler:2007,Szalay:2008,Splotch:2010}, and see 
\citet{hassan:2010} for a solution using a distributed cluster with GPUs; and using the dataset characteristics or optimized data structure to 
provide a multi-resolution visualization solution \citep{hopf:2003,hopf:2004}.

Of all the applications of scientific visualization in astronomy that we have examined, $N$-particle data provides the closest match to, and hence 
may be the greatest beneficiary of, advances in the wider field of scientific visualization.  Their particular use of scattered and grid 
data formats means that general purpose visualization packages (see Section \ref{sct:software}) are more suitable for handling simulation data,
notwithstanding the need to convert from custom astronomy data formats to required input formats. There may be some benefit in providing simple file 
conversion tools, otherwise astronomers using simulation data may care to investigate alternative standard data representations (e.g. HDF5\footnote{http://www.hdfgroup.org/HDF5/}, VTK file format). 
While sharing some similarities with gridded simulation data, visualization of spectral data cubes presents some unique problems, which we now explore.

\begin{table*}
\caption{The classification of contributions dedicated to the visualization of
spectral data cubes. See Section \ref{sct:outline} and \ref{sct:DataNature} for a description of the visualization techniques and dataset types. The methodology column highlights the main implementation trend used in each work.  }
\label{tab:ListofCubePapers}
\centering
\begin{tabular}{|p{5.5cm}|c|c|c|c|c|c|}
\hline
& \multicolumn{2}{|c|}{Dataset Type} & \multirow{2}{*}{Methodology} & \multicolumn{3}{|c|}{Visualization Techniques}\\ \cline{2-3} \cline{5-7}
&Radio&IFU&&Isosurface&VR&Slicing\\ \hline
\citet{domik:1992a,brugel:1993}&$\bullet$&&Modify&$\bullet$&$\bullet$&$\bullet$\\ \hline
\citet{gooch:1995b,gooch:1996,oosterloo:1995, oosterloo:1996}&$\bullet$&&Custom&&$\bullet$&\\ \hline
\citet{plante:1999}&$\bullet$&&Modify&&$\bullet$&\\ \hline
\citet{beeson:2003}&$\bullet$&&Modify&&$\bullet$&\\ \hline
\citet{miller:2006}&&$\bullet$&Existing&&$\bullet$&\\ \hline
\citet{borkin:2005,borkin:2007}&$\bullet$&&Existing&$\bullet$&$\bullet$&$\bullet$\\ \hline
\citet{li:2008}&$\bullet$&$\bullet$&Custom&&$\bullet$&\\ \hline		
\citet{draper:2008}&$\bullet$&&Modify&$\bullet$&$\bullet$&\\ \hline
\citet{hassan:2010}&$\bullet$&&Custom&&$\bullet$&\\ \hline	
\end{tabular}
\end{table*}

\subsection{Spectral Data Cubes}
\label{sct:SDC}

A spectral data cube comprises two spatial dimensions (usually RA and Dec, or galactic longitude and latitude), one frequency or wavelength dimension, 
and a flux value.\footnote{This may be any one of the four Stokes parameters, but is usually Stokes I.}  The frequency or wavelength dimension is often 
converted to a line-of-sight velocity using Doppler-shift relationships.  
Spectral data cubes are more common in radio astronomy, however, with improvements in integral field units (IFUs) and similar instrumentation, optical/IR astronomy is seeing a growth in the collection and use of spectral cubes.

Spectral data cubes are characterized by a lack of well defined surfaces, low signal to noise data values combined with a high dynamic data range, 
and the use of special coordinate systems that do not always match well with the equal-unit 3D spatial coordinates of other disciplines.
This limits the use of existing general purpose scientific visualization tools, particular in comparison to simulation data. It is still very common for astronomers to rely on 2D techniques, such as data slicing or projected moment maps, as the primary method for visualizing data.  These approaches can be achieved using packages like SAOImage DS9\footnote{\url{http://hea-www.harvard.edu/RD/ds9/}}, or some modules of Karma \citep{gooch:1996}. However, with terabyte-scale data cubes from near term facilities telescopes like ALMA, ASKAP, and MeerKAT, slicing techniques may not be feasible (e.g. it would take ~30 minutes to step through a ~16,000-channel ASKAP cube at 10 frames/sec, assuming that there was software capable of supporting this approach), and the complex kinematics of, e.g. an unresolved pair of merging galaxies, may not be fully captured using a 2D moment map. 

Early work by \citet{norris:1994}, \citet{gooch:1995b}, and \citet{oosterloo:1995} emphasised the opportunity for 3D visualization to aid in the processes of data analysis and knowledge discovery.  Since then, three main techniques have been used to visualize spectral data cubes: isosurfacing, volume rendering, and (2D) data slicing.  Volume rendering has dominated 3D spectral cube visualization in general due to its ability to give the user a global data perspective, despite the lack of well defined surfaces within the observational data, and its improved capability of visualizing in the low 
signal to noise regime.  

Table \ref{tab:ListofCubePapers} shows the key publications relating to spectral data cube visualization. Several comments need to be made on the table:
\begin{itemize}
	\item The data source classification was based on the experimental data presented in each paper.
	\item The implementation trend ``Modify'' is used to describe the published work if it uses an external library or tool to perform/support the visualization. In some cases [e.g. \citet{beeson:2003}] the application may add a lot of customizations and code enhancement to effectively handle the astronomical datasets. 
	\item It is clear that the number of publications in this research branch is relatively fewer than the publications in $N$-particle visualization. 
	\item As noted before volume rendering is the key visualization technique and is supported by all the tools.
\end{itemize}

The implementation trends for spectral cube visualization are similar to those found for particle data:
\begin{itemize}
	\item	Modify/extend an existing general purpose visualization library [e.g. \citet{domik:1992a,brugel:1993,plante:1999, beeson:2003, draper:2008}]. 
	\item	The use of some visualization packages developed to serve the medical visualization domain such as Paraview, 3D slicer, and Osirix \citep{borkin:2007}.
	\item	Build a custom visualization library/application \citep{gooch:1996,kahler:2007,li:2008,hassan:2010}.
\end{itemize}
 
Special attention has been given to VTK\footnote{\url{http://www.vtk.org/}} as a basis for modifications/extensions achieved via the first implementation trend \citep{plante:1999,draper:2008}. On the other hand, it is also possible to use some of the packages classified as $N$-particle visualization tools for spectral data. Although no explicit examples have been shown through their publications, most of the $N$-particle tools capable of handling structured grids can visualize spectral data cubes [e.g. \citet{gheller:2003, amati:2003,becciani:2010, kahler:2007}].

While the usage of high performance computing architecture and hardware play an important role in visualizing $N$-particle datasets, only \citet{beeson:2003}, \citet{li:2008} and \citet{hassan:2010} use that approach to improve the rendering speed or to handle larger than memory spectral datasets. 

We believe 3D spectral data cube visualization is still in its infancy. There is a need to move spectral data cubes visualization tool from ``pretty picture''-generating tools into powerful data analysis tools. Spectral data visualization tools should provide their users with: quantitative visualization capabilities, ability to handle huge datasets exceeding single machine processing capacity, accepted interactivity levels, and effective noise suppression techniques. Also, offering powerful two-way integration with other data analysis and reduction tools will be key to facilitate the wider usage of such tools. We will further discuss these issues within Section \ref{sct:discuss}.

We now turn our attention to other developments in scientific visualization that are not based on the data representation: distributed and remote
visualization services, collaborative visualization, visualization workflows, and public outreach outcomes. With the exception of public outreach 
visualization, there has been much less research effort expended in these areas for astronomy, and consequently, their level of community up-take is 
somewhat lower than for the particle and spectral cube approaches we have examined.

\subsection{Distributed/Remote Visualization}
\label{sct:dv}
Desktop computers have a finite memory size, typically a few gigabytes, yet many datasets from observation and simulation are much larger than this 
(e.g. processed ASKAP data cubes will be over 1 terabyte).  A solution to 
this problem lies in the use of distributed visualization, where a networked computing cluster shares the processing tasks.  

A typical astronomer does not always have immediate access to sufficient computational power or data storage capacity to deal effectively with such large datasets. Moreover, effective and efficient implementation of software to deal with large datasets requires a higher-level of computing knowledge 
relating to choice and use of appropriate data structures, techniques for
scheduling, etc. 
Remote visualization therefore presents an opportunity to provide the wider astronomy community with a visualization service with potentially lower cost, administrative effort, and a reduced need to transfer data. At the same time it presents a cost effective way to further utilize existing expensive computational infrastructure. This philosophy was the main motivation for the Virtual Observatory (VO) concept of sharing datasets, and providing astronomers with data analysis and visualization software as a service [see \citet{Quinn:2004}, and \citet{Williams:2009} for details]. Rather than requiring local hardware, a user requests a visualization of a dataset from a remote host - the outcome of the visualization, usually an image, is returned to the user.  Along with the time taken to produce images, such a system has an overhead in terms of the interaction speed and the network speed. 

The issue of providing 3D visualization and computational infrastructure as a service was addressed by \citet{plante:1999}, \citet{murphy:2006}, 
and \citet{becciani:2010}. All of them agreed on using the Web as the main service platform. \citet{plante:1999} built a custom VRML viewer using Java3D to render the output of a VTK-based server. The custom VRML viewer enable them to provide additional interactivity services such as a 3D cursor, the ability to select subregions, and produce 2D JPEG snapshots of the visualization output. The same methodology was used by \citet{beeson:2004} to visualize data from catalogue streamed in XML format, but with a ready made browser plugin to render VRML output.   
\citet{murphy:2006} describe an image display remote visualization service (RVS)\footnote{\url{http://www.atnf.csiro.au/vo/rvs/}} through a set of VO tools for the storage, processing and visualization of Australia Telescope Compact Array data. The RVS server accepts FITS images and provides a 2D visualization using an AIPS++ back end.
 
The last and probably one of the most complete systems is VisIVO web\footnote{\url{http://visivoweb.oact.inaf.it/visivoweb/index.php}} \citep{becciani:2010}. The system uses Web 2.0 technology for user interaction, while the output is generated as static images, with a semi-interactive control of dataset orientation and movie generation. It is a simple way to implement such functionality, but is perhaps not as intuitive as the interaction provided through custom web controls or environments such as Java3D.

The usage of distributed processing to enable astronomers to handle larger than memory datasets was addressed by \citet{beeson:2003}, \citet{Splotch:2010}, 
and \citet{hassan:2010}. \citet{beeson:2003} extended the shear warp volume-rendering algorithm \citep{lacroute:1994} with a distributed implementation. 
Demonstrated using both spectral line data cubes and $N$-body datasets, their technique relies on distributing the volume data among the participating computing nodes and then using the associative ``over'' operator to yield a final image. Their code was based on Virtual Reality Volume Rendering (VIRVO) code\footnote{\url{http://www.calit2.net/~jschulze/projects/vox/release/deskvox2_00b.txt}}. 

\citet{Splotch:2010} developed a custom ray tracing code to render, in a fast and effective way, point like data. They exploit the use of multi-core architecture using OpenMP \citep{splotch:2008}, distributed memory architecture using MPI, and GPUs using the CUDA development toolkit. Their technique was demonstrated using $N$-body datasets only. \citet{hassan:2010} used a distributed GPU cluster to enable ray-casting volume rendering of datasets up to  26 Gbytes at frame rates better than 5 frames/sec. Combining shared and distributed memory high performance computing capabilities enabled them to handle large than memory datasets at an accepted frame rate with a lower number of nodes than \citet{beeson:2003}. 


\subsection{Collaborative Visualization}
\label{sct:cv}
Collaborative visualization enables multiple users to share 
a visualization experience.  For this to be successul, the main requirements are
high-speed networks and effective communication protocols.
Early work in this field was by \citet{van:1995}, who  implemented
the AstroVR Collaboratory environment for distributed users to share in 
the analysis of FITS images.  Communication between users was 
via audio, video, or typed text.  The AstroVR approach was motivated by
early client-server, multi-user networked games\footnote{Also known 
as Multi-User Dungeons or MUDs.}. \citet{plante:1999}  described
collaborative support in the NCSA Horizon Image Data Browser (Version 2.0) 
via NCSA Habanero\footnote{\url{http://www.isrl.illinois.edu/isaac/Habanero/}}.  
Collaborators were able to join a {\em hablet}  
session following an e-mail invitation, with interaction
via the GUI visible to all participants.

Both \citet{djorgovski:2009} and \citet{nakasone:2009} consider the use of
virtual environments based on the Second 
Life\footnote{\url{http://secondlife.com}} framework developed by Linden
Labs. Launched in 2003, Second Life is an online,  
multi-user, virtual world application with support for real-time interaction,
creation and exploration of three-dimensional environments, and synchronous 
communication (including both text and voice). As of early 2010, Second 
Life had more than 16 million registered accounts, although $\sim 40,000$ 
``residents'' are typically online at any one time.  It is the collaborative 
experience that is of most interest to astronomers - geographically 
distributed users can interact simultaneously with a data visualization, with 
feedback on what the other participants are doing/seeing.   The 
main drawback at present is the limited support for large astronomical 
datasets: the Second Life application imposes a limit of 15,000 objects, 
leaving  \citet{nakasone:2009} to experiment 
with visualizing only 1024 particles from a stellar cluster simulation.
\citet{djorgovski:2009} propose adapting 
OpenSim\footnote{\url{http://opensimulator.org}},
an open source implementation of Second Life, which may partly remove
that restriction. 


\subsection{Image Display and Interaction}
\label{sct:image}
While not unique to scientific visualization, the use of advanced displays for
presentation of, and interaction with, three-dimensional datasets is worthy of
consideration.   Advanced displays may include tiled or multi-display 
walls,  stereoscopic environments
ranging from flat-screens to immersive CAVE-like environments, and domes
(upright and tilted). 

Early descriptions of the limitations of 2D displays for 3D astronomical data are in
\citet{fomalont:1982} and \citet{rots:1986}. \citet{rots:1986} described the use of a mosaic
of 2D slices, time-sequence animations,  the creation of 3D solid surfaces
(ie. an isosurface at a given threshold level), and the possibilities
offered by stereoscopic images and holograms!
\citet{fluke:2006} considered a suite of advanced displays including
multi-panel or tiled displays, digital domes, and stereoscopic projection, 
with descriptions of low-cost implementations of each display.  
Comprehensive overviews of stereoscopic and 3D display systems and 
technologies may be found in \citet{mcallister:2006} or \citet{holliman:2006}.

One of the main challenges is the lack of native support for advanced displays 
from visualization software \citep{fluke:2006}.  Most advanced displays require 
images in a different format to conventional displays (i.e. on a monitor or 
data projector). This includes fish-eye or other spherical projection for 
domes, and image pairs for stereoscopic displays.  There is an overhead 
in producing such frames, which can have a negative impact on frame rates and
hence usability.  On the other hand, viewing data with an advanced display 
may yield additional insight.  Apart from a few projects using CAVE-like
environments, stereoscopic and dome display has mostly been reserved
for public outreach visualization (see Section \ref{sct:public}). 

A related issue, which has yet to achieve a satisfactory resolution, is the 
choice of appropriate 3D interaction device. Intuitive and easy real-time 
interaction with visualization output, including changing visualization 
parameters, camera position, and interactive dynamic data filtering \citep{Shneiderman:1994}, is vital to achieve the required visualization 
outcomes. This may also be one of the reasons why quantitative techniques
in 3D have not advanced \citep{gooch:1995a,gooch:1995b}, 
as they rely on a device (e.g. for selection of objects or regions) that is 
as simple to use in 3D as the mouse is for interacting with 2D data.  

Few astronomy publications have explictly addressed practical 
alternatives for interacting with astronomical data.  \citet{gooch:1995a} 
considered the 6-degree of freedom Spaceball (Spatial Systems, Inc.) as 
an alternative to manipulate a 3D cursor within a 3D volume. The Spaceball 
was a low-cost version of the devices used in immersive environments, 
controlling additional functionality such as interactive slicing. 
\citet{kahler:2002}, and \citet{kahler:2002IEEE} used a voice and gesture controlled 
CAVE application to define a camera path following the interesting features. 


\subsection{Workflow}
\label{sct:workflow}
When dealing with a large amount of datasets, additional benefits may be achieved using workflow driven applications. Selecting a certain visualization parameter is not usually a straight forward process: knowledge about the visualization algorithms, and the dataset characteristics are essential to achieve sensible visualization outcomes. Indeed, reproducing specific visualization results is 
challenging, particular when an interactive process has been used to control
properties such as data limits, transparency, colour maps and orientation.
Keeping metadata about the visualization process itself through an integrated workflow management system was addressed by \citet{kapferer:2008}. They discussed the usage of VisTrials\footnote{\url{http://www.vistrails.org/index.php/Main Page}} as a scientific workflow management system that provides support for data exploration and visualization.

On the other hand, providing the user with visualization software that tightly integrates with simulation, data analysis, or data generation tools may facilitate the use of new visualization tools/techniques, remove the data conversion barrier, and provide a better interoperability. Some published work [e.g. \citet{dubinski:2008, comparato:2007}] discusses this concept, and different real time data sharing and integration protocols were introduced as a part of the VO initiative [e.g \citet{taylor:2006}].   

\citet{li:2008} presents a visualization workflow for multi-wavelength 
astronomical data. The importance of their work comes from the completeness of 
their proposed framework that included GPU-based data processing and new ways 
to visualize multi-wavelength astronomical data with volume visualizations 
(such as the ``horseshore'' model). They offer a collection of interactive 
exploration tools tailored for multi-wavelength datasets.

\subsection{Public Outreach}
\label{sct:public}

Astronomy has a demonstrated history of engaging public interest in 
science - this has been achieved in large part by the appealing visuals 
that are routinely generated from telescopes and simulations.  
As a professional astronomer, there is something special about being able 
to share the results of our research work with the public.  It is easy 
for our passion to inspire audiences of all ages, from the youngest 
school students to adults.  However, the techniques that we often use 
in collecting, understanding and exploring our datasets (histograms, 
scatter plots, error-bars, etc.) do not always make for visually appealing, 
or necessarily understandable, presentations. Public outreach visuals are qualitative or occasionally comparative, 
but rarely quantitative.

In some sense, public outreach use of scientific visualization techniques
has exceeded that for science outcomes, with a number of advancements 
in astronomy visualization motivated by outreach or presentation goals.  
At times, the divide between
an outreach visualization, and one that is intended to help astronomers
to gain deeper understanding of their models and observerations, is very
narrow.  Cases in point are the highly realistic renderings of
the Orion nebulae \citep{emmart:2000}, emission nebulae, and planetary nebulae \citep{magnor:2004}, which we describe below, and the previous discussion of AMR 
visualizations starting with \citet{kahler:2002IEEE} (section \ref{sct:largeN}).

In 2000, within the SIGGRAPH 2000 electronic theatre, a team from the San
Diego Supercomputing Center (SDSC) presented a volume rendering video 
animation for the Orion nebula \citep{emmart:2000}. \citet{nadeau:2001} and \citet{genetti:2002} described this 
work in more detail, including their use of a volume scene
graph \citep{nadeau:2000}. They reported a set of limitation in 
the available volume rendering applications/libraries, including a lack
of efficient parallel perspective volume rendering, forcing them to build
a  customized parallel perspective viewing model, and the need for high voxel
resolution to capture all details over a wide range of
length scales (from proplyds within the central region of the nebula, 
with a scale of 0.007 light years, to the outskirts of the nebula
at 14.3 light years).   Additionally, to achieve a sufficient level
of photorealism with regards to the glowing gas in the nebula, 
existing treatment of transparency as the inverse of the opacity did not
work -  it was necessary to modify the modelling and rendering tools
to allow independent values for transparency and opacity.

Realistic planetary nebula models were created using constrained inverse 
volume rendering by \citet{magnor:2004}.  As a purely emissive model 
(i.e. glowing gas), the fast,
texture-based volume rendering technqiue of \citet{cabral:1994} was found
to be an appropriate visualization solution. The goal of this work was to create
realistic looking planetary nebula, enabling models to be fitted to 
three observed systems with bipolar symmetry. \citet{magnor:2005} implemented a solution to the more complex case of generating interactive 
volume renderings of 3D models with dust (e.g. reflection nebulae) by 
using GPU-based ray-casting.  A scattering depth is assigned to each voxel 
of the nebulae, and these are accumulated along the view-dependent line-of-sight.
The code was able to handle multiple illuminating stars along with
multiple scattering events.

Computer techniques have greatly simplified the process of creating custom animations.  
Starting with segments like {\em Where the Galaxies Are} \citep{geller:1992} using
data from the CfA redshift surveys, the ``galaxy distribution fly-through'' 
has become a standard way to visualize galaxy redshift survey data.

Public outreach visualization often requires the combination of disparate 
data sets and data types -- and visualization packages.  
For example, the stereoscopic movie {\em Cosmic Cookery} \citep{holliman:2006}, 
used data from the 2dF Galaxy Redshift Survey \citep{colless:2001}, large-scale structure from the
Millennium simulation \citep{springel:2005a}, higher-resolution galaxy formation 
simulations, and conceptual animations to link between sequences.   Packages
used for this movie included Celestia,\footnote{\url{http://www.shatters.net/celestia}}
PartiView, 3dsMax, VolView by 
KitWare,\footnote{\url{http://www.kitware.com/products/volview.html}}
and custom rendering software. 

{\em ``Solar Journey''}, implemented in VRML and OpenGL, demonstrates components of
the solar envirnoment as both an interactive environment and a short animated 
film \citep{hanson:2002}.  
Problems that needed to be solved included registration of multispectral
datasets, texture-mapping of objects from Earth-based images, and difficulities
with providing a completely accurate spatial model when true spatial information 
on some features was limited.

While opportunities for further public outreach using 3D visualization exist, the
reality is that productions of this nature come at a cost.  They are
time-consuming to produce, which is not necessarily an incentive for 
researchers who are already time poor.  They usually require access to 
(sometimes expensive) commercial animation packages (e.g.
Maya, Lightwave, 3DSMax) and experienced animators, both of which rarely can be justified within research-only budgets.  
These same animation packages are not designed for the types of data 
that astronomers use, so there a need for significant conversions of 
datasets to animation-friendly format (e.g. FITS Liberator).  Creating 
flight-paths can be a non-trivial process.  Rendering, 
the process of creating individual frames that are then brought together 
to form a movie sequence, can take from minutes to hours per frame, and 
can require access to a supercomputer or dedicated render farm.

The challenge is to simplify the process of creating engaging 
visualisations that can also enhance research productivity.  That 
is, visualisation tools that are not just designed for public presentation, 
but can also be used in the academic context for conference presentations, 
research publications, or department/personal web pages.  Astronomers usually 
do not need (or want) pre-rendered computer animations to analyze
their data - real-time, interactive solutions are much more valuable.

\section{Visualization Software}
\label{sct:software}

While astronomers have written about their own visualization software, there is no summary of the variety of packages that are available. \citet{leech:2005} surveyed visualization software suitable for sub-mm spectral line data, considering user requirements such as FITS format,  
compatability with astronomical coordinates, support for mosaicking, display of 2D slices and moment maps, and quantitative capabilities.  3D rendering
functionality was considered ``a plus''. They weighed up the pros and cons of AIPS++ \citep{mcmullin:2004}, the Starlink \citep{draper:2005} Software
Collection (specifically the {\tt kappa} and {\tt datacube} packages, which mostly offer visualization tasks via the command line), Karma [\citet{gooch:1996} - {\tt kviz} offers 2D slicing, while {\tt xray} is a volume rendering packages],
OpenDX\footnote{\url{http://www.opendx.org}} (a general purpose visualization package based on IBM's Visual Data Explorer), although data conversion to the OpenDX {\tt .dx} file format was required, the 
PDL\footnote{\url{http://pdl.perl.org}} Perl module which supports FITS and NDF formats, and Python using the PyFITS
module \citep{barrett:2000}.  The main conclusion of the comparison was that ``no single software package met all of the user requirements'', many tools
lacked a GUI, there were opportunities for comparing 3D software across wavelengths to avoid re-developement, and open-source licensing was desirable. See \citet{perez:1997} for a related comparison of the issues facing astronomers when choosing between custom astronomical software and 
commercial packages for data analysis. 

\subsection{Custom code}
\label{sct:Customcode}
While there has been much effort to date in creating general purpose visualization tools (such as Paraview\footnote{\url{http://www.paraview.org/}}, 
VisIT\footnote{\url{https://wci.llnl.gov/codes/visit/}} 
and AMIRA\footnote{\url{http://www.amira.com/}}), many of these existing 
software packages are not suitable for astronomy due to:
\begin{itemize}
\item limitations with handling specific astronomy data formats [e.g. FITS or the GADGET-2 file format] which require a data format conversion process before using these tools [e.g fits2itk\footnote{\url{http://am.iic.harvard.edu/FITS-reader}}]. This data format conversion disables the direct real-time integration and may imply increase in the dataset size;
\item the need for conversion from astronomically meaningful units (RA, Dec, redshift) to general units (cm or mm in three-dimensions), which often limits the user to exploring data in a qualitative form only [see \url{http://am.iic.harvard.edu/UsingSlicer} for an example];
\item the high-dynamic range low signal-to-noise domain in which many observational projects work; and
\item the data volumes (billion particle data generated for high-resolution cosmological simulations or many giga-voxels for high resolution spectral cubes).
\end{itemize}
These issues necessitate the creation of domain-specific applications, and solving visualization problems that are unique to astronomy. However, utilizing existing visualization general purpose libraries is a good starting point.  

In Table \ref{tab:ListofPack1} and \ref{tab:ListofPack2}, we provide a list of libraries and packages aimed at supporting scientific visualization of astronomical data.  This list does not make any claims on completeness or suitability of apackage 
for a particular dataset. Web links were correct (and live) at the time of writing. VTK and OpenGL are the main workhorses for scientific visualization, providing the basis for many of the listed tools, however, as programming 
and development environments, they have a reasonably steep learning curve, and what may seem like simple tasks can take some time to code.  The advantage of a pre-existing visualization package or library (either general purpose or astronomy-focused) is that someone else has dealt with implementation issues, which should mean that you 
can get to a science outcome faster.  The down-side is that any pre-existing package may not be able to do {\em exactly} what you require it to do.

\onecolumn

\begin{longtable}{|c|c|c|p{3.5cm}|c|p{5cm}|}
\caption{List of the supported visualization techniques for some of the most popular visualization packages used in astronomy}
\label{tab:ListofPack1} \\
\hline
\multirow{2}{*}{Package} & \multicolumn{2}{|c|}{Rendering} & \multirow{2}{*}{Techniques} & \multirow{2}{*}{Last Update} & \multirow{2}{*}{Website}\\ \cline{2-3} 
&2D&3D&&&\\ \hline
\multirow{3}{*}{3D Slicer}&&\multirow{3}{*}{$\bullet$}&Volume Rendering&\multirow{3}{*}{2010}&\multirow{3}{*}{\url{http://www.slicer.org/}}	\\
&&& Isosurface && \\
&&& Label Map && \\ \hline
\multirow{3}{*}{AIPS++/CASA}&\multirow{3}{*}{$\bullet$}&\multirow{3}{*}{}&Raster&\multirow{3}{*}{2010}&\url{http://casa.nrao.edu/}\\ 
&&& 2D Contouring && \multirow{2}{*}{\url{http://www.astron.nl/aips++}}\\ 
&&& Vector && \\ \hline
Amira&$\bullet$&$\bullet$&Most Volume, Surface, Scatter visualization&2010&\url{http://www.amira.com/}																																	\\ \hline
\multirow{3}{*}{AstroMD}&\multirow{3}{*}{$\bullet$}&\multirow{3}{*}{$\bullet$}&Scatter Plot&\multirow{3}{*}{2004}&\multirow{3}{*}{\url{http://cosmolab.cineca.it/}} \\
&&& Isosurface   && \\
&&& Volume Rendering  && \\ \hline
DVR&&$\bullet$&Volume Rendering&2004&{Not Available Online}	\\ \hline
\multirow{2}{*}{Glnemo}&\multirow{2}{*}{$\bullet$}&\multirow{2}{*}{$\bullet$}&3D scatter Plot&\multirow{2}{*}{2009}&\multirow{2}{*}{\url{http://www.oamp.fr/dynamique/jcl/glnemo}} \\
&&& 2D contouring&&											\\ \hline
Glnemo2&$\bullet$&$\bullet$&3D Scatter Plot&2010&\url{http://bima.astro.umd.edu/nemo/man_html/glnemo2.1.html}  \\ \hline
GNUPlot&$\bullet$&$\bullet$&Scatter Plot&2010&\url{http://www.gnuplot.info/}  \\ \hline
Hubble in a Bottle&&$\bullet$&Volume Rendering&2007&\url{http://hubble.sourceforge.net/}																						\\ \hline
IDL&$\bullet$&$\bullet$& Most Volume, Surface, Scatter visualization&2009&\url{http://www.ittvis.com/ProductServices/IDL.aspx}																					\\ \hline
\multirow{4}{*}{IFRIT}&\multirow{4}{*}{$\bullet$}&\multirow{4}{*}{$\bullet$}&Volume Rendering&\multirow{4}{*}{2010}&\multirow{4}{*}{\url{http://sites.google.com/site/ifrithome/}} \\
&&& Stream Tube && \\
&&& Isosurface && \\
&&& 2D Contouring &&	\\ \hline
\multirow{3}{*}{Karma}&\multirow{3}{*}{$\bullet$}&\multirow{3}{*}{$\bullet$}&Raster&\multirow{3}{*}{2006}&\multirow{3}{*}{\url{http://www.atnf.csiro.au/computing/software/karma/}}\\
&&& Volume Rendering && \\
&&& 2D Contouring &&		\\ \hline
OpenDX&$\bullet$&$\bullet$&Most Volume, Surface, Scatter visualization&2007&\url{http://www.opendx.org/index2.php}																											\\ \hline
\multirow{2}{*}{Osirix}&&\multirow{2}{*}{$\bullet$}&Volume Rendering&\multirow{2}{*}{2010}&\multirow{2}{*}{\url{http://www.osirix-viewer.com/}} \\
&&& IsoSurface &&																												\\ \hline
Paraview&$\bullet$&$\bullet$&Most Volume, Surface, Scatter visualization&2010&\url{http://www.paraview.org/}																														\\ \hline
PartiView&$\bullet$&$\bullet$&Scatter Plot&2007&\url{http://www.haydenplanetarium.org/universe/partiview}																\\ \hline
\multirow{2}{*}{RVS}&\multirow{2}{*}{$\bullet$}&&Raster&\multirow{2}{*}{2005}&\multirow{2}{*}{\url{http://www.atnf.csiro.au/vo/rvs}}	\\
&&&2D Contouring &&\\ \hline
\multirow{4}{*}{S2Plot}&\multirow{4}{*}{$\bullet$}&\multirow{3}{*}{$\bullet$}&Volume Rendering &\multirow{4}{*}{2009} &\multirow{4}{*}{\url{http://astronomy.swin.edu.au/s2plot/index.php?title=S2PLOT}} \\
&&& Isosurface && \\
&&& Vector Map && \\
&&& 2D Contouring &&														\\ \hline
\multirow{3}{*}{SPLASH}&&\multirow{3}{*}{$\bullet$}&Volume Rendering&\multirow{3}{*}{2010}&\multirow{3}{*}{\url{http://users.monash.edu.au/~dprice/splash/}} \\
&&& Vector Plot&& \\
&&& 2D contouring &&																						\\ \hline
StarSplatter&&$\bullet$&Scatter Plot&2007&\url{http://www.psc.edu/Packages/StarSplatter_Home/}																	\\ \hline
\multirow{2}{*}{TIPSY}&&\multirow{2}{*}{$\bullet$}&Scatter Plot&\multirow{2}{*}{2009}&\multirow{2}{*}{\url{http://hpcc.astro.washington.edu/tools/tipsy/tipsy.html}} \\
&&& 2D contouring &&																\\ \hline
\multirow{2}{*}{TopCat}&\multirow{2}{*}{$\bullet$}&\multirow{2}{*}{$\bullet$}&Scatter Plot&\multirow{2}{*}{2010}&\multirow{2}{*}{\url{http://www.star.bris.ac.uk/~mbt/topcat/}} \\
&&& Line/Spherical Plot &&																							\\ \hline																													
\multirow{4}{*}{VisIVO}&&\multirow{4}{*}{$\bullet$}&Scatter Plot&\multirow{4}{*}{2010}&\multirow{4}{*}{\url{http://visivo.oact.inaf.it/index.php}} \\
&&& Isosurface&& \\
&&& Volume Rendering && \\
&&& 2D Contouring &&		\\ \hline																													
\multirow{3}{*}{VOPlot3D}&\multirow{3}{*}{$\bullet$}&\multirow{3}{*}{$\bullet$} &Scatter Plot&\multirow{3}{*}{2009}&\url{http://vo.iucaa.ernet.in/~voi/VOPlot3D_UserGuide_1_0.htm} \\
&&& Surface Plot &&\\
&&& Histogram &&	\\ \hline
\end{longtable}


\begin{longtable}{|c|c|c|c|c|c|}
\caption{List of the supported data representation for some of the most popular visualization packages used in astronomy}
\label{tab:ListofPack2} \\ \hline
\multirow{2}{*}{Package} & \multicolumn{5}{|c|}{Data Representation}\\ \cline{2-6} 
&Point-Like&Structured Grid&3D Cube&Images&Catalogues\\ \hline

3D Slicer&&$\bullet$&$\bullet$&$\bullet$&	\\ \hline
AIPS++/CASA&&&&$\bullet$& \\\hline
Amira&$\bullet$&$\bullet$&$\bullet$&$\bullet$&$\bullet$			\\ \hline
AstroMD&$\bullet$&$\bullet$&&$\bullet$&$\bullet$ \\ \hline
DVR&&$\bullet$&$\bullet$&&	\\ \hline
Glnemo&$\bullet$&&&&	\\ \hline
Glnemo2&$\bullet$&&&&	\\ \hline
GNUPlot&$\bullet$&&&& \\ \hline
Hubble in a Bottle&$\bullet$&&&&$\bullet$	\\ \hline
IDL&$\bullet$&$\bullet$&$\bullet$ &$\bullet$&	$\bullet$				\\ \hline
IFRIT&$\bullet$&$\bullet$&$\bullet$&& $\bullet$\\ \hline
Karma&&$\bullet$&$\bullet$&$\bullet$&		\\ \hline
OpenDX&$\bullet$&$\bullet$&$\bullet$&$\bullet$&	$\bullet$	\\ \hline
Osirix&&$\bullet$&$\bullet$&& \\ \hline
Paraview&$\bullet$&$\bullet$&$\bullet$&$\bullet$&$\bullet$	\\ \hline
PartiView&$\bullet$&&&$\bullet$&				\\ \hline
RVS&&&$\bullet$&$\bullet$& \\ \hline
S2Plot&$\bullet$&$\bullet$&$\bullet$ &$\bullet$ &$\bullet$ \\ \hline
SPLASH&$\bullet$&$\bullet$&&& \\ \hline
StarSplatter&$\bullet$&&&&							\\ \hline
TIPSY&$\bullet$&&&&										\\ \hline
TopCat&$\bullet$&&&&$\bullet$			\\ \hline																								
VisIVO&$\bullet$&$\bullet$&$\bullet$&&$\bullet$ \\ \hline							
VOPlot3D&$\bullet$& &&&$\bullet$ \\ \hline

\end{longtable}

\twocolumn

\section{Discussion}
\label{sct:discuss}
In this review of scientific visualization in astronomy, we have 
attempted to provide an overview of the work that has been undertaken
over the last two decades.  A few key techniques from the broader
discipline of scientific visualization have been adopted by astronomers,
most notably the use of volume rendering and scattered point representations,
while others are rarely used (in particular, streamlines and vector 
visualization).  
Based on our assessment of the literature, we now consider some of the key challenges for wider adoption and
research into relevant scientific visualization techniques for astronomy
from the viewpoints of visualization researchers and astronomers.

\subsection{Challenges for Visualization Researchers}
Although the format of astronomy datasets may be familiar to people working in scientific 
visualization, astrophysical datasets have a set of special characteristics that may 
limit the usability of some more general visualization techniques despite them having a much wider user-base and higher level of technical support. 
These features can be summarized into the following points:
\begin{enumerate}
\item {\em Lack of dominant efficient data representation}.  Within some astronomy sub-fields, there is no single dominant data representation. For example, $N$-body simulation data can exists in different data formats such as the Gadget-2 file format, and custom ASCII or binary formats. Also, there is a lack of standard data representation for catalogues. 
Although the introduction of the VOTable format (proposed by the International Virtual Observatory Alliance) represents an attempt to unify data in Internet-accessible format, it is not yet widely used.  
Not only does this raise interoperability issues but also makes the development of a generic astronomy visualization packages either complex or incomplete. On the other hand, some current commonly-used astronomy data representations (especially FITS) are more oriented toward data archiving rather than efficient data accessibility. This limits the application's ability (not only visualization applications) to provide users with fast data loading, disables the usage of out-of-core algorithms, and disables the usage of distributed data storage. An exception to this is the usage of NCSA hierarchical data form (HDF)\footnote{\url{http://www.hdfgroup.org}} which permits parallel I/O and enables distributed data storage \citep{ostriker:1997}.

Different packages solve this problem by using a single data format for its internal implementation and provide users with an importing functionality that converts existing commonly-used data formats into the internal file format [e.g \citet{sanchez:2004,sanchez2004a, kissler:2004, becciani:2010}]. This may not be an efficient solution for large datasets due to its storage and processing requirements.

\item {\em Low signal to noise ratio and high dynamic range.} Data generated from radio/optical telescopes often combines a low signal to noise ratio with a large dynamic range. This requires special data manipulation and interpolation schemas, which may reduce its effect on the final resultant visual output.

\item {\em Use dimensions in a different way.} Most of the current visualization algorithms and applications are designed to visualize data assigned to 2D/3D spatial grids (e.g. CFD and medical grids). Grids (if they exist) in astrophysical datasets may contain different dimension types (i.e: redshift) in combination with the regular spatial domains. In some cases the dimensional information is mentioned as a type of metadata, so the visualization algorithm must first make a correct mapping between 
the data axes and the data values to be capable to use current known visualization algorithms. This also limits the usage of  general purpose visualization packages in a quantitative manner.  

\item {\em Huge datasets:} As noted in the previous section, data volumes from
large-$N$ particle simulations and high-resolution spectral data cubes
routinely exceed millions (and often billions) of data points.  These 
present problems relating to the memory and computational demands to handle
such data sizes; the need to support high levels of interactivity,
such as shifting quickly through different spatial scales (Brunner et al.
2002); or streaming of such data volumes.

\end{enumerate}

\subsection{Challenges for Astronomers}
\label{sct:discuss2}
Scientific data visualization can, and does, provide opportunities to support a wide range of existing and 
planned astronomy research projects. Through our investigation of the literature, we have determined the following reasons why scientific visualization techniques may not have achieved a more widespread usage in astronomy:

\begin{enumerate}

\item {\em The lack of quantitative tools that integrate seamlessly with visualization.} This issue was identified
by \citet{norris:1994}, but little progress has been made to address it.  Using annotations and providing the user with  quantitative information about their data, combined with the qualitative visualization output, is an essential add-on to facilitate the data analysis and exploration task. Only a few of the published astronomy visualization works show the need for interactive and quantitative visualization and provide a proposed solution for that [e.g. \citet{amati:2003}, \citet{ahrens:2006}, and \citet{li:2008}].

\item {\em Visualization is not science for the astronomers.} The success of an astronomical 
project is judged by the science result it produces. The time invested by an astronomer in 
becoming an expert in using or developing visualization software must be balanced 
against the expected scientific gain. It is difficult to justify and obtain funding based 
purely on methodological approaches such as visualization and data mining, even if such an 
approach will demonstrably improve the scientific return \citep{ball:2009} . 
We think that this may be the main reason why most of the astronomy visualization 
trials neither lasted for a long period nor resulted in a widely used application.

\item {\em Visualization does not do the science.} The successful interpretation of 
visualization result is up to the scientist. The output may not 
represent a straightforward relationship or pattern. Visualization researchers 
aim to simplify this interpretation step, but adding a meaning to the final 
visualization output is still the astronomer's task.

\item {\em Adjustable parameters and technique configuration.} Selecting a suitable iso-value, 
a colour map, or a transfer function is not always directly related to the 
dataset type or format. Sometimes this may require a deep understanding of both the data and the 
visualization technique. The focus of many projects was to visualize a certain object or was with a limited objective. Only a small number of projects 
were targeting general datasets or community usage of the software produced [e.g. \citet{becciani:2000}, \citet{barnes:2006}, and \citet{becciani:2010}]. 

\item {\em Usability and interoperability.} Only the applications with focused visualization 
functionality, easy to use user interfaces, easy to deploy instructions, and dedication to a certain data type were widely used by the astronomical 
society. Also being a cross platform application is an important consideration for usability. 

\item {\em There is a problem in citing the astronomy  visualization effort or 
application}.  In preparing this review, we found it very difficult to
determine an approximate number of users for each package.  Quite often,
astronomy-focused visualization software is not supported by an obvious,
citable research paper, so there is a missed opportunity for developers to
receive tangible credit for software that is being used to help support 
research.  In some cases, we suspect that the lack of on-going development 
or support for some applications may be tied to what is slowly being recognized
as a wider issue for software developers in astronomy \citep{weiner:2009}.

\item{\em Where do scientific visualization papers get published?}  The papers 
contributing to our review come from a range of astronomy journals, conference proceedings, and non-astronomy journals.   The full selection of papers does not appear in ADS searching, meaning that astronomers may not 
be aware of their existence - particularly if the trend of not citing
visualization papers continues.  Our work serves a purpose
in highlighting some of the more important and relevant papers in the field.

\end{enumerate}

\subsection{Six Grand Challenges for the Petascale Astronomy Era}

We assert that visualization has a critical role to play in maximizing 
the scientific return of astronomical data in the Petascale Astronomy Era. However,
to achieve this goal, work is required to overcome the following challenges:
\begin{enumerate}

\item {\em Support Quantitative Visualization.} 

To advance astronomy visualization tools from  ``pretty picture'' generating tools into effective knowledge discovery tools, astronomers need integrated quantitative support in addition to the currently provided qualitative output. Although the current qualitative data views are vital to give astronomers global pictures of their data, and have the potential to play an increased role in 
quality control of data (see below), there is a need to extend astronomy visualization tools to better support ``doing'' science. 
The ability to apply different data filters, inspect data points/objects for certain properties, overlay external catalogues/maps, apply mathematical operations,  and select different sub-regions for further study are examples of missing tools. Most of these tools exist in current two-dimensional data analysis and processing packages (such as Karma) but these offer limited support for three-dimensional data. 

Although providing such tools may seem easy, large data sizes and low signal-to-noise properties will be a limitation for any effective implementation. Also, interaction with three-dimensional data will need a major change to enable effective implementation of such functionality (see below).

\item {\em Effective Handling of Large Data Sizes.}

Handling petabyte datasets will be a big challenge for most astronomical data analysis packages. Real-time (or near real-time)  interaction requirements worsen the situation for astronomy visualization. Additionally, storage requirements, networking costs, and transfer speeds will force many astronomers to change their way of doing science. 

While most of the development effort now is going towards automated data analysis and information extraction systems, it is noteworthy that:
\begin{itemize}
\item There is no automated system that reaches a 100\% recognition rate. Specifically with the low signal-to-noise data that is regularly used for knowledge discovery, it is computationally challenging to achieve a high rate of automated pattern recognition. 
\item Scientific visualization is expected to play an important role as a quality control tool for the output from different automated tools and data reduction processes. Keeping all the raw data for some new astronomy instruments (e.g ASKAP and MWA) will not be feasible, so the trend will be to overwrite this data after the data reduction process is done.  In such cases, the usage of scientific data visualization will be vital to detect and overcome any defect in the data gathering and early analysis process.
\end{itemize}

\citet{hassan:2010} addressed some of these issues and provided a solution for visualizing larger-than-memory datasets with the aid of distributed processing and GPUs. However, improving data formats to support distributed storage and parallel file accessing, extending current data analysis tools to deal effectively with larger-than-memory datasets, and extending/enhancing  current data analysis tools with techniques more suitable to 3D datasets are still missing steps in the astronomical data processing pipeline.

\item {\em Discovery in  Low Signal-to-noise data.}

Effectively dealing with noise is an important aspect when dealing with astronomy data.  With large data sizes, the low signal to noise properties of astronomy datasets limit the usage of multi-resolution techniques and out of core methodologies.  On the other hand, new discoveries usually happen near the noise level.  Here, visualization plays a vital role in the absence of effective automated tools. 
Although noise suppression and removal is a signal processing problem and may fall outside the scope of scientific visualization, advanced usage of colors to enhance comprehension and the development of customized transfer functions and shaders should be a priority for the next generation of astronomy visualization tools.

The use of colors  in order to enhance comprehension has received little attention in 
astronomy. With the notable exception of \citet{rector:2005,rector:2007} there has been little work in investigating whether
an application of colour science and visual grammar (such as composition,
orientation, etc.) can improve cognition. 
This should include a detailed investigation of the benefits and limitations 
of using different colour maps and colour contrasts [such as harmonic
colourmaps - \citet{wang:2008}].  Some effort has been made in the use of colour information to 
produce photorealistic rendering of simulation data, such as in the 
AMR work on first stars by \citet{kaehler:2006}, however, this is mostly 
with a public outreach product in mind.  In the case of spectral data from a 
radio telescope, there is no such property as ``photorealism'', and a 
pseudo colour map must be used for intensity or fractional polarization.   

Also, the use of transfer functions (or color mapping schemas) to provide astronomers with better insight into their data and suppressing noise will boost the effectiveness of visualization usage as a qualitative data analyses technique [see  \citep{gooch:1995a,gooch:1995b} for customized transfer functions developed for radio astronomy]. 
There does not seem to have been any systematic investigation of the use of 
shaders in astronomy visualization, yet in other application domains, this is one of the main areas of research
interest [see \citet{sato:1998,sato:2000}, \citet{li:2003}, and \citet{correa:2008} for example].

\item {\em Better Human-Computer Interaction and Ubiquitous computing.}  

The standard 
computer mouse is well-suited for interaction with two-dimensional data,
but is limited when moving to multi-dimensional data. Immersive VR environments provide alternative
interaction, often through a ``wand'', which gives the user better control 
of navigation in a three-dimensional world, but this can require an expensive infrastructure or increased computational resources to drive. We believe that using relatively cheaper game controllers (e.g the Nintendo Wii controller), or programmable handheld interaction devices (specifically the Apple iPad) as tool for exploration and navigation within advanced visualization environments, will be a better and affordable alternative. Such low cost and easy-to-use solutions may help to bring advanced interaction to the astronomer's desktop.    
Also, using ubiquitous computing methodologies to make interaction with visualization output more accessible and easier for astronomers may increase the adoption of 3D visualization [see \citet{greensky:2008} for example of using such techniques in other scientific domains].

\item {\em Better Workflow Integration. }

Improved integration between different data analysis and visualization tools will encourage the usage of scientific visualization. A typical example is providing two-way interaction between source finders and 3D visualization tools, where the output of the source finder is overlaid on the visualization output. synthesis between data analysis/processing and visualization applications will provide the user with better control over, and give more insight into, the data analysis processes, and potentially speed up the knowledge discovery process.   

Seamless astronomy\footnote{\url{http://research.microsoft.com/en-us/events/ersymposium2009/seamless_astronomy.pdf}}, a concept advocated by 
Alyssa Goodman 
and colleagues, is an example which takes workflow issues one step further [see \citet{Goodman:2009yz} 
for a discussion on the role of `modular craftmanship' in data visualization].  
Here, complete integration between different areas of the typical astronomer's 
workflow, such as linking searches for astronomical objects with related 
research papers, are all wrapped up within an interactive desktop 
environment.  This is an extension of a service-oriented computing 
paradigm for astronomers, where the connectivity between applications is 
more natural.  While early experimentation has centred on the World-Wide 
Telescope\footnote{\url{http://www.worldwidetelescope.org}} application,
such an approach could incorporate 3D visualization either through support
of remote visualization services, or cross-platform browser-based applications.

\item {\em Encourage adoption of 3D scientific visualization techniques.} 

As we discussed in section \ref{sct:discuss2}, there has been a somewhat limited effort by the astronomical community to develop widely-used, general purpose, visualization tools targeted at astronomical data.  The lack of financial investment, compared to other disciplines which are motivated by both financial and
social outcomes, limits the effort towards a complete astronomy visualization product with a suitable high level of documentation and support. 

Furthermore, there appears to be a social issue limiting the usage of current available tools. This may be due to the learning curve required to master a particular application, or the perceived low level of scientific return from the amount of time invested in using an alternative visualization technique. The lack of technical support, easy-to-use tools, and complete documentation contribute to limiting the amount of users’ benefits from existing tools.  

We expect that this may change during the next decade with the increasing demand for better knowledge discovery and data analysis pipelines, capable of dealing with the upcoming data avalanche.  Building national/community facilities will provide a solution for that on the short-term. Extending the use of remote-visualization and data analysis tools through the web-platform will minimize the time and effort required to adopt visualization and offers a cost effective solution by minimizing the total cost of ownership and reduces the need to transfer upcoming huge datasets.

\end{enumerate}

\section{Conclusion}
\label{sct:conclusion}

We have investigated the state of scientific visualization research as
applied to the domain of astronomy. As with the overall discipline, the use of 
computer-based scientific visualization in astronomy is still a young field. 
Over the last two decades, the astronomy community has invested substantially 
in domain-specific 2D and 3D visualization tools and techniques. For 
example, ``Astronomical Data Analysis Software \& Systems'' (ADASS)
the major international annual astronomy software conference, has hosted 
more than 30 presentations on astronomy data visualization during this 
period. However, few research groups have concentrated for extended periods 
of time on improving visualization techniques for astronomy, and 
consequently, the potential for new visualization approaches to improve 
scientific output in the discipline has not been thoroughly explored.  

Scientific visualization is a genuine research technique in its own right and is a 
fundamental, enabling technology for knowledge discovery. Great discoveries in astronomy are not always related to the patterns 
we already know -- it is related to the identification of strange phenomena, unexpected relations, 
or unknown patterns.  These cannot be achieved with an automated tool. There is still much
that can be learnt about the process of scientific visualization and its domain
specific application in astronomy. With the Petascale Astronomy Era on the
horizon, there has never been a more pressing need.



\section*{Acknowledgments} 
This research was supported under Australian Research Council's
Discovery Projects funding scheme (project number DP0665574). We have benefited at various stages from disussions with David Barnes 
(Swinburne), and Alyssa Goodman and members of the IIC (Harvard).
We thank Madhura Killedar (University of Sydney) for allowing us to 
visualize her cosmological dataset.This research has made extensive use of 
NASA's Astrophysical Data System Bibliographic services.

\bibliographystyle{elsarticle-harv}        
\bibliography{references}           

\end{document}